\documentclass[twocolumn,showpacs,floats,floatfix,superscriptaddress,aps,pra]{revtex4-1}

\usepackage{amsfonts}
\usepackage{amssymb}
\usepackage{amsmath}
\usepackage{calc}
\usepackage{dcolumn}
\usepackage{graphicx}
\usepackage{bm}
\usepackage{epstopdf}
\usepackage{epsfig}
\usepackage{float}
\usepackage {color}
\usepackage{textcomp}
\usepackage{soul}
\usepackage{diagbox}

\begin{document}

\title{Arbitrary polarization control by magnetic field variation}

\author{Elena Stoyanova}
\affiliation{Department of Physics, Sofia University, James Bourchier 5 blvd, 1164 Sofia, Bulgaria}

\author{Svetoslav Ivanov}
\affiliation{Department of Physics, Sofia University, James Bourchier 5 blvd, 1164 Sofia, Bulgaria}

\author{Andon Rangelov}
\affiliation{Department of Physics, Sofia University, James Bourchier 5 blvd, 1164 Sofia, Bulgaria}

\begin{abstract}
We propose a universal scheme for the construction of a device for arbitrary to arbitrary polarization transformation, which consists of two quarter-wave plates and two Faraday rotators. Using this device, one can continuously change the retardance and the rotation angle simply by changing the magnetic fields in each Faraday rotator.
\end{abstract}

\maketitle

\def\blue{}
\def\brown{}
\def\cyan{}
\def\yellow{}
\def\green{}
\def\red{}
\def\black{}


\section{Introduction}

In many practical applications the ability to observe and control polarization is critical,
as polarization is one of the basic characteristics of transverse light waves \cite%
{Hecht,Wolf,Azzam,Goldstein,Duarte}. Two prominent optical devices for controlling light's polarization state are optical
polarization rotators and optical polarization retarders \cite%
{Hecht,Wolf,Azzam,Goldstein,Duarte}. A polarization rotator rotates the plane
of linear light polarization at a specified angle \cite{Pye,Damask,Rangelov,Dimova,Stojanova},
while a retarder (or a waveplate) introduces a phase difference between two orthogonal
polarization components of a light wave \cite{Pye,Damask,Ardavan,Ivanov,Peters}.

Retarders are usually made from birefringent materials. Fresnel rhombs \cite{Bennett,Bakhouche} are also widely used
and, based on total internal reflections, they achieve retardation at a broader range of wavelengths.
Two common types of retarders are the half-wave plate
and the quarter-wave plate. By introducing a phase shift of $\pi$ between the two orthogonal
polarization components for a particular wavelength, the half-wave plate effectively rotates the polarization vector
to a predefined angle.
The quarter-wave plate introduces a shift of $\pi/2$, and thereby converts linearly polarized light into circularly
polarized light and vice versa \cite{Hecht,Wolf,Azzam,Goldstein,Duarte}.

Although half-wave and quarter-wave plates clearly dominate in practice,
retarders of \emph{any} desired retardation can be designed and successfully applied.
Tuning the retardance is a valuable feature because in some practical settings one may need a half-wave plate,
while in others a quarter-wave plate is required.
Tunable retardance can be achieved by liquid-crystals \cite{Sharp,Sit,Ye}.
Alternatively, one can use the technique recently suggested by Messaadi et al. \cite{Messaadi},
which is based on two half-wave plates cascading between two quarter-wave
plates. This basic optical system functions as an adjustable retarder that
can be controlled by spinning one of the half-wave plates.

A polarization rotator may employ Faraday rotation or birefringence. The Faraday
rotator consists of a magnetoactive material that is put inside a powerful
magnet \cite{Hecht,Wolf,Azzam,Goldstein,Duarte}. The magnetic field causes a
circular anisotropy (Faraday effect), which makes left- and right-circular
polarized waves ``feel'' different refraction indices.
As a result, the linear polarization plane is rotated. A birefringent rotator
can be achieved as a combination of two half-wave plates. Such a rotator would
have an angle of rotation equal to twice the angle between the optical axes of
the two half-wave plates \cite{Zhan}.

Wave plates and rotators are basic building blocks for polarization
manipulation. Indeed, every reversible polarization transformation,
(a reversible change in the polarization vector from any initial state to any final state)
can be achieved using a composition of a retarder and a rotator \cite{Hurvitz}.
An arbitrary transformation requires one half-wave plate and two quarter-wave plates \cite{Simon},
or just two quarter-wave plates \cite{Bagini,Zela,Damask} if the
apparatus itself is rotated. Arbitrary polarization transformations, however,
require one to perform rotations on individual plates which may turn out to be very
impractical in particular applications where one needs to change the angles with certain frequency and speed.

In this paper we attempt to solve this problem by substituting mechanical rotations of the plates with
variation of magnetic fields.
The device we propose is a modified version of Simon-Mukunda's controller and consists of two
quarter-wave plates and two rotators. In this setting we can perform fast and continuous variation
of the polarization vector simply by changing the magnetic fields in each Faraday rotator.


\section{Prefaces}


The Jones matrix that describes a rotator with rotation angle $\theta$ is
\begin{equation}
\mathbf{R}(\theta )=\left[
\begin{array}{cc}
\cos \theta & \sin \theta \\
-\sin \theta & \cos \theta%
\end{array}%
\right] ,
\end{equation}%
while the Jones matrix that represents a retarder is
\begin{equation}
\mathbf{J}(\varphi )=\left[
\begin{array}{cc}
e^{i\varphi /2} & 0 \\
0 & e^{-i\varphi /2}%
\end{array}%
\right].
\end{equation}%
Here $\varphi $ is the phase shift between the two orthogonal
polarization components of the light wave. The most widely-used retarders are the half-wave plate ($%
\varphi =\pi $) and the quarter-wave plate ($\varphi =\pi /2$) \cite%
{Pye,Damask}.

\begin{figure}[htb]
\centerline{\includegraphics[width=0.8\columnwidth]{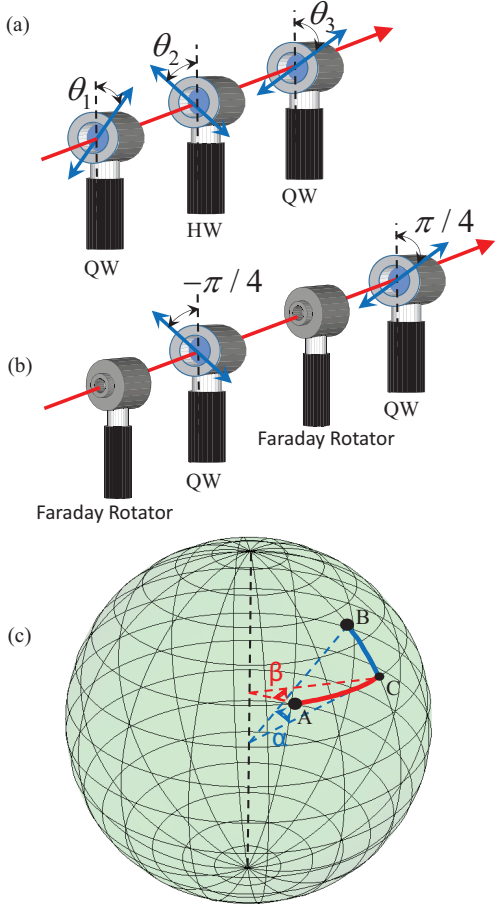}}
\caption{(Color online)(a) The Simon--Mukunda polarization controller in
a configuration of the form QW-HW-QW (b) The scheme of arbitrary to arbitrary
polarization transformation device, composed by two quarter-wave plates and
two Faraday rotators. The orientation of the quarter-wave plates is fixed.
(c) Polarization evolution on the Poincare sphere. The initial polarization is at
point A and the final polarization is at point B. As can be seen from Eq. (\protect\ref%
{arbitrary-to-arbitrary polarization}) the first part of the evolution is
rotation at angle $\protect\beta$ between points A and C, followed by a retardation
at angle $\protect\alpha$ from point C to point B.}
\label{fig1}
\end{figure}

Consider now a single polarizing birefringent plate of phase shift $\varphi$, whose fast
axis is rotated to an angle of $\theta$ relative to the vertical axis (azimuth angle).
In a 2-dimensional rectangular coordinate system whose axes are aligned with the horizontal
and the vertical directions, the Jones matrix is given by
\begin{equation}
\mathbf{J}_{\theta }(\varphi )=\mathbf{R}(-\theta )\mathbf{J}(\varphi )%
\mathbf{R}(\theta ).  \label{retarder}
\end{equation}
If behind this plate one places another plate with azimuth angle $\theta +\alpha /2$,
the resulting Jones matrix is given by the product
\begin{equation}
\mathbf{J}_{\theta +\alpha /2}(\pi )\mathbf{J}_{\theta}(\pi )=-\left[
\begin{array}{cc}
\cos \alpha & -\sin \alpha \\
\sin \alpha & \cos \alpha%
\end{array}%
\right] ,  \label{rotator}
\end{equation}%
that represents a Jones rotator matrix (up to an unimportant phase of $\pi$) \cite{Zhan,Rangelov}.


\section{Modified Simon--Mukunda polarization controller}


A general scheme of a device capable of arbitrary polarization
transformations is the Simon--Mukunda
polarization controller \cite{Simon}. It consists of one half-wave plate
(HW) and two quarter-wave plates (QW) in one of the following arrangements
QW-HW-QW, QW-QW-HW, or HW-QW-QW. The first one is the most popular (Fig. \ref{fig1} a)
and is also used as a fiber polarization controller \cite{Thorlabs}. The
Simon--Mukunda polarization controller \cite{Simon} operates in the following way: the
first quarter-wave plate turns the input elliptical polarization into a linear
polarization. Then the half-wave plate rotates the obtained linear polarization vector,
which is finally transformed into the required elliptical polarization output by the second quarter-wave plate.
The Jones matrix for the Simon--Mukunda polarization controller is
\begin{equation}
\mathbf{J}=\mathbf{J}_{\theta _{3}}(\pi /2)\mathbf{J}_{\theta _{2}}(\pi )%
\mathbf{J}_{\theta _{1}}(\pi /2).
\end{equation}%
Because the unit matrix can be written as
\begin{equation}
\mathbf{\hat{1}}=\mathbf{J}_{\theta _{1}}(\pi /2)\mathbf{J}_{\theta
_{1}}(-\pi /2),
\end{equation}%
we obtain
\begin{equation}
\mathbf{J=J}_{\theta _{3}}(\pi /2)\mathbf{J}_{\theta _{2}}(\pi )\mathbf{J}%
_{\theta _{1}}(\pi /2)\mathbf{J}_{\theta _{1}}(\pi /2)\mathbf{J}_{\theta
_{1}}(-\pi /2).  \label{polarization controller}
\end{equation}%
Next we use that
\begin{equation}
\mathbf{J}_{\theta _{1}}(\pi )\mathbf{=J}_{\theta _{1}}(\pi /2)\mathbf{J}%
_{\theta _{1}}(\pi /2)
\end{equation}%
to simplify Eq. (\ref{polarization controller}):
\begin{equation}
\mathbf{J=J}_{\theta _{3}}(\pi /2)\mathbf{J}_{\theta _{2}}(\pi )\mathbf{J}%
_{\theta _{1}}(\pi )\mathbf{J}_{\theta _{1}}(-\pi /2).
\end{equation}%
Next we make use of Eq. (\ref{rotator}) for the combination of two half-wave
plates
\begin{equation}
\mathbf{R}(2(\theta _{1}-\theta _{2}))=-\mathbf{J}_{\theta _{2}}(\pi )\mathbf{%
J}_{\theta _{1}}(\pi ),
\end{equation}%
to get the final expression of the Jones matrix:
\begin{equation}
\mathbf{J}=-\mathbf{J}_{\theta _{3}}(\pi /2)\mathbf{R}(\alpha )\mathbf{J}%
_{\theta _{1}}(-\pi /2),  \label{Simon--Mukunda}
\end{equation}%
where the rotator angle is $\alpha =2(\theta _{1}-\theta _{2})$. Therefore the
Simon--Mukunda polarization controller can be constructed as combination of
two quarter-wave plates along with a rotator between them. This device would operate
in a similar way as the one before: the first quarter-wave plate turns the
input elliptical polarization into a linear polarization vector, which is then rotated
by the rotator element, and is finally transformed into the required elliptical output
polarization by the second quarter-wave plate.


\section{Arbitrary retarder as a special case of the modified Simon--Mukunda
polarization controller}


In the special case when the two quarter-wave plates are oriented such that
their fast optical axes are perpendicular to each other (cf. Eq. \ref%
{Simon--Mukunda}) ($\theta _{1}=\theta _{3}=\pi /4$) we obtain a retarder
with retardation $2\alpha $ \cite{Messaadi}:
\begin{equation}
\mathbf{J}_{0}(2\alpha )=\mathbf{J}_{\pi /4}(\pi /2)\mathbf{R}(\alpha )%
\mathbf{J}_{\pi /4}(-\pi /2)=-\left[
\begin{array}{cc}
e^{i\alpha } & 0 \\
0 & e^{-i\alpha }%
\end{array}%
\right] .  \label{tunable retarder}
\end{equation}%
If the two quarter-wave plates are achromatic (for example as in\textbf{\ }%
\cite{Ardavan,Ivanov,Peters} or if Fresnel rhombs are used as quarter-wave
plates) then one can achieve a wavelength tunable half-wave or quarter-wave plate that
operates differently compared to previously suggested tunable wave plates \cite%
{Goltser,Darsht}. However, in contrast to previously used tunable wave plates,
here we do not need to rotate the wave plates, but rather change the
magnetic field in the Faraday rotator. \red Therefore, the
suggested tunable retarder is not mechanical and can be used as a fast
switcher, where the switching on/off time of the optical activity is in the
order of microseconds with the state of the art approaching subnanosecond
\cite{Didosyan,Shaoying}. \black

\section{Arbitrary to arbitrary polarization converter}


Based on the fact that combining an arbitrary rotator with an arbitrary
retarder allows to achieve any polarization transformation \cite{Hurvitz},
we can combine the tunable retarder from Eq. (\ref{tunable retarder}) with
an additional rotator to get an arbitrary-to-arbitrary polarization
manipulation device. Its Jones matrix is
\begin{equation}
\mathbf{J}=\mathbf{J}_{0}(2\alpha )\mathbf{R}(\beta ),
\end{equation}%
or
\begin{equation}
\mathbf{J}=\mathbf{J}_{\pi /4}(\pi /2)\mathbf{R}(\alpha )\mathbf{J}_{\pi
/4}(-\pi /2)\mathbf{R}(\beta ).  \label{arbitrary-to-arbitrary polarization}
\end{equation}%
%
The proposed optical device, illustrated schematically in Fig. \ref{fig1} b, has potential
advantages over the Simon--Mukunda polarization controller
\cite{Simon,Damask}, where one has to adjust the spatial orientation of all three wave plates.
The proposed device (cf. Eq. (\ref{arbitrary-to-arbitrary polarization})) is more convenient to use
in the sense that the rotator angle and the retardation are obtained by changing the magnetic field of the first
and the second Faraday rotator, respectively. Furthermore, our scheme
is fast switchable on and off, because in the absence of a magnetic field the
polarization is not changed.

\begin{figure}[tbh]
\centerline{\includegraphics[width=0.8\columnwidth]{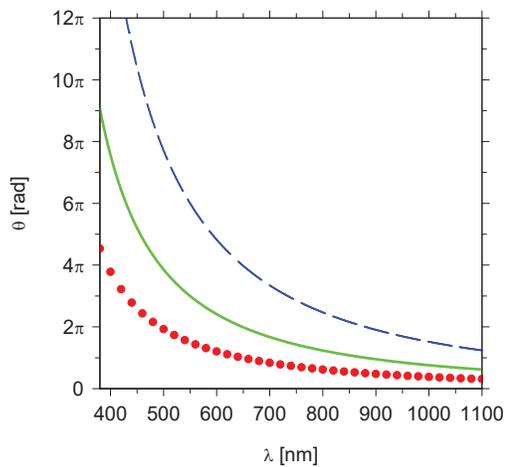}}
\caption{\red (Color online) The Faraday rotation angle $\protect\theta $ vs the
light wavelength $\protect\lambda $ for three different magnetic fields $%
B_{1}=0.5$ T (red dotted), $B_{2}=1$ T (green solid) and $B_{3}=2$ T (blue
dashed). \black}
\label{fig2}
\end{figure}

\red
Finally, we investigate the possibility of achieving any pair of rotation angles $%
\alpha $ and $\beta $ (cf. \eqref{arbitrary-to-arbitrary polarization}) for
the proposed arbitrary-to-arbitrary polarization conversion device.
The rotation angle for the Faraday rotator is given by%
\[
\theta (\lambda )=V(\lambda )BL\,,
\]%
where $B$ is the external magnetic field, $L$ is the magneto-optical element
length, and $V(\lambda )$ is Verdet's constant. We do the
most common calculation involving the Terbium Gallium Garnet (TGG) crystal, as this crystal yields a
high Verdet constant. So far, the dispersion of Verdet's constant for the TGG crystal
has been extensively studied \cite{Bozinis1978, Jannin1998,
Yoshida,Villora2011} and the wavelength dependence has been
shown to be described by the formula
\begin{equation}
V(\lambda )=\frac{E}{\lambda _{0}^{2}-\lambda ^{2}}\,,  \label{dispersion}
\end{equation}%
where $E=4.45\cdot 10^{7}\,\frac{\text{rad $\cdot $ nm}^{2}}{\text{T $%
\cdot $ m}}$ and $\lambda _{0}=257.5$~nm is the wavelength, often close to
the Terbium ion's 4f-5d transition wavelength. In the range of 400 --- 1100 nm,
excluding 470 --- 500 nm (absorption window \cite{Villora2011}), the TGG crystal has
optimal material properties for a Faraday rotator. The Verdet constant
decreases with increasing wavelength for most materials (in absolute value):
for the TGG crystal it is equal to $475\,\frac{\text{rad}}{\text{T $\cdot $ m}}$ at $400$%
~nm and $41\,\frac{\text{rad}}{\text{T $\cdot $ m}}$ at $1064$~nm \cite%
{Villora2011}. Our simulations were carried out for three different values
for the magnetic field, $B_{1}=0.5$ T, $B_{2}=1$ T and $B_{3}=2$ T, at a fixed length $%
L=0.05$ m of the TGG crystal. As can be seen from Fig. \ref{fig2}, any pair of
rotation angles $\alpha $ and $\beta $ in the interval $[0,2\pi ]$\ can be achieved with magnetic field smaller than 1T for the visible spectrum.
Therefore, with commercial Faraday rotators available on the market, the
practical realization of the proposed polarization control device should be
straightforward.

\black

\section{Conclusion}


In conclusion, we have suggested two useful polarization manipulation
devices. The first device is the modified Simon--Mukunda polarization
controller, which in contrast to the traditional controller
is constructed as a combination of two quarter-wave plates and a Faraday
rotator between them. Our second device for arbitrary to arbitrary
polarization transformation is composed of two Faraday rotators and two
quarter-wave plates, where the retardance and the rotation can be continuously
modified merely by changing the magnetic fields of the two Faraday rotators.
Because they use Faraday rotation, the suggested schemes are non-reciprocal.
We hope the proposed methods for polarization control would be cost-effective and useful in any scientific laboratory.


\section*{Acknowledgment}

This work was supported by Sofia University Grant 80-10-191/2020.

\end{document}